\documentstyle{article}
\begin{document}
\hoffset -1.5cm

\title{\centerline{\Large\bf IS U(1)$_H$ A GOOD FAMILY SYMMETRY?}}

\author{\large William A. Ponce$^{1,2}$, Arnulfo Zepeda$^2$ and
Jes\'us M. Mira$^1$\\
\normalsize 1-Departamento de F\'\i sica, Universidad de Antioquia \\
\normalsize A.A. 1226, Medell\'\i n, Colombia.\\
\normalsize 2-Departamento de F\'\i sica,
Centro de Investigaci\'on y de Estudios Avanzados del IPN.\\
\normalsize Apartado Postal 14-740, 07000 M\'exico D.F., M\'exico.\\}

\maketitle
\vspace{2.cm}

\renewcommand{\baselinestretch}{1.2}
\setlength{\hsize}{15cm}
\setlength{\vsize}{10in}

\hspace{.2cm}

\large

\begin{center}
{\bf ABSTRACT}
\end{center}

\parbox{15cm}%
{We analyze U(1)$_H$ as a horizontal symmetry and its possibilities to explain
the known elementary-fermion masses. We find that only two candidates, in the
context of
SU(3)$_c\otimes$SU(2)$_L\otimes$U(1)$_Y\otimes$U(1)$_H$
nonsupersymmetric, are able to fit the experimental result m$_b<<$m$_t$.}

\vfill
\noindent
Accepted for publication in Z. Phys. C
\pagebreak

\section{INTRODUCTION}
The pattern of fermion masses, their mixing, and the family replication,
remain as the most outstanding problems of nowadays particle physics. The
successful standard model (SM) based on the local gauge group
SU(3)$_c\otimes$SU(2)$_L\otimes$U(1)$_Y$ can
tolerate, but not explain the experimental results. Two main features that a
consistent family theory should provide are:\\
(i)-Within each charge sector, the masses increase with generation by large
factors:
\begin{center}
$m_u<<m_c<<m_t; \hspace{.5cm} m_d<<m_s<<m_b; \hspace{.5cm}
m_e<<m_\mu<<m_\tau$
\end{center}
(ii)-Even if one restricts to the heaviest family, the masses are still quite
different:
\begin{center}
$m_\tau\sim m_b<<m_t.$
\end{center}
The horizontal survival hypothesis\cite{hsh} was invented in order to
accommodate (i), under the (wrong) assumption that $m_\tau\sim m_b\sim
m_t$. The idea of radiative symmetry breaking in a supersymmetric extension of
the SM\cite{dssm} depends crucially on the existence of one quark
with a mass comparable to the SM breaking scale, but it can not explain why
this was the top quark instead of the bottom quark. The modified horizontal
survival hypothesis\cite{mhsh} was introduced in order to explain the full
extent of (i) and (ii), but a dynamically realization of this hypothesis is
still lacking. Of course, these hypothesis and ideas rest on the assumption
that all the dimension four Yukawa couplings in a well behaved theory should
be of order one.

Related to (i) and (ii) is the fact that the Cabibbo-Kobayashi-Maskawa quark
mixing matrix is near to the identity, but it is a common prejudice to
assume that
the appropriate family symmetry may explain this fact as a consequence of (i)
and (ii). In what follows we will enlarge the SM gauge group with an extra
U(1)$_H$ horizontal local gauge symmetry (the simplest multi-generational
continuous symmetry we can think of). We then show that the
structure SU(3)$_c\otimes$SU(2)$_L\otimes$U(1)$_Y\otimes$U(1)$_H$ by itself
is able to explain (ii), and that the simplest supersymmetric (SUSY) extension
of this model without a $\mu$-term can not cope with (ii).

\section{SU(3)$_c\otimes$SU(2)$_L\otimes$U(1)$_Y\otimes$U(1)$_H$  as an
anomaly-free model}
Our attempt is to keep the number of assumptions and parameters down to the
minimum possible, and to try to construct a model which explains both
features (i)
and (ii) at the lowest possible energy scale. We therefore
demand cancellation of the triangular (chiral) anomalies\cite{anom} by the
power counting method, including the mixed gravitational(grav)
anomaly\cite{del}.
The alternative of cancelling the anomalies by a Green-Schwarz
mechanism\cite{gs} has been already considered in Refs.\cite{papa},
and corresponds to the construction of a model string-motivated which demands
the inclusion of physics near the Plank scale.

SU(3)$_c\otimes$SU(2)$_L\otimes$U(1)$_Y\otimes$U(1)$_H$  as a continuous
gauge group, with U(1)$_H$ as a family symmetry, was introduced
long ago in Ref.\cite{koca}, (revived recently in the context of SUSY
string-motivated models in Refs.\cite{papa,shrock,nir}).
There are two different
versions of the model, corresponding to two different ways of cancelling the
chiral anomalies. One is demanding cancellation of the anomalies
for each generation and the other one is cancelling the anomalies between
generations.

\subsection{Cancellation of anomalies in each generation}

Assuming there are no right-handed neutrinos, using the U(1)$_Y$ and
U(1)$_H$ charges displayed in Table (1), and demanding freedom from chiral
anomalies for\\
SU(3)$_c\otimes$SU(2)$_L\otimes$U(1)$_Y\otimes$U(1)$_H$, we get:

\begin{eqnarray}
{[{\rm SU}(2)_L]^2U(1)_H }&:& Y_{\psi_i}+3Y_{\chi_i}=0\\
{[{\rm SU}(3)_c]^2U(1)_H} &:& 2Y_{\chi_i} + Y_{U_i}+Y_{D_i}=0\\
{[U(1)_Y]^2U(1)_H }&:& 2Y_{\psi_i}+4Y_{E_i}+\frac{2}{3}Y_{\chi_i} +
\frac{16}{3}Y_{U_i} + \frac{4}{3}Y_{D_i} = 0\\
U(1)_Y[U(1)_H]^2 &:& -Y^2_{\psi_i}+Y^2_{E_i}+Y^2_{\chi_i}
-2Y^2_{U_i}+Y^2_{D_i}=0\\
{[{\rm grav}]^2U(1)_H }&:& 2Y_{\psi_i} + Y_{E_i} = 0\\
{[U(1)_H]^3 }&:& 2Y^3_{\psi_i} + Y^3_{E_i} + 6Y^3_{\chi_i}
+ 3Y^3_{U_i}+3Y^3_{D_i} = 0.
\end{eqnarray}
The solution to Eqs.(1)-(6) is \cite{koca,po}

\[ Y^i_{H_\eta}=\alpha_iY_{SM_\eta}, \]

\noindent
where $\alpha_i$ is an arbitrary number
different for each generation, and Y$_{SM_\eta}$ is the U(1)$_Y$ charge
for the $\eta$ multiplet.

These U(1)$_H$ charges cannot explain the feature (ii) which demands that
at tree level only the top quark acquires a mass, and therefore that the
Higgs field with U(1)$_H$ charge Y$_{H\phi}$ satisfies:

\[ Y_{\chi_3}+Y_{U_3}=Y_{H\phi} \]

\[ Y_{\chi_3}+Y_{D_3}\neq -Y_{H\phi}=Y_{H\phi^\star}. \]

\noindent
But once the first of these equations is satisfied, Eq. (2) above implies
$Y_{\chi_3}+Y_{D_3}=-Y_{H\phi}$. Therefore, if a top quark mass arises at
tree level ($Y_{H\phi}=\alpha_3$), a bottom mass arises as well at the
same level.

Adding right-handed neutrinos N$^c{i,_L}$ to our set of fundamental
fields does not change this conclusion since Eq.(2) stays valid [the
only changes are in Eqs. (5) and (6) which are now replaced by
\begin{eqnarray}
{[{\rm grav}]^2U(1)_H} &:& \hspace{.5cm} 2Y_{\psi_i}+Y_{E_i}+Y_{N_i}=0
\\ \nonumber
{[U(1)_H]^3} &:&
2Y^3_{\psi_i}+Y^3_{E_i}+6Y^3_{\chi_i}+3Y^3_{U_i}+3Y^3_{D_i}+Y^3_{N_i}=0].
\nonumber
\end{eqnarray}

\subsection{Cancellation of anomalies between generations}
If the U(1)$_H$ anomalies are cancelled by an interplay among
generations, Eqs. $(1)-(6)$ should be understood with a summ over
$i=1,2,3$. Eq. (4) then reads

\begin{equation}
\sum_i(-Y^2_{\psi_i}+Y^2_{E_i}+Y^2_{\chi_i}-2Y^2_{U_i}+Y^2_{D_i})=0.
\end{equation}

\noindent
Obviously a solution to the new anomaly constraint equations which are
linear or cubic in the Y$_{\eta_i}$ is

\[ \sum_{i=1}^3 Y^i_{\eta}=0 \]

\noindent
for each $\eta$. We will limit ourselves to this type of solutions and
within this set we will consider only those for which the $\psi_i$ and
$U_i$ H-hypercharges are fixed to satisfy either

\[Y_{\psi_1}=\delta_1\equiv\delta,Y_{\psi_2}=\delta_2=-\delta,Y_{\psi_3}
=\delta_3=0,\]

\[Y_{U_1}=\delta_1^\prime\equiv\delta^\prime,Y_{U_2}=\delta_2^\prime=-
\delta^\prime,Y_{U_3}=\delta^\prime_3=0, \]

\noindent
or any set of relations obtained from the former equations by a
permutation of the indices $i=1,2,3$. The solutions can then be divided
onto four classes according to the way the cancellations occur in Eq.(8).

\noindent
\underline{CLASS A}\\
$Y_{E_i}=Y_{\psi_i}=\delta_i$ and $Y_{D_i}=Y_{\chi_i}=Y_{U_i}=\delta_i^\prime;
i=1,2,3$. A model with a tree-level top quark mass arises if
$Y_{H_\phi}=Y_{\chi_i}+Y_{U_j}$ for some $i$ and $j$. There are
five different models in this class characterized by
$Y_{H_\phi}=\pm 2\delta^\prime,\pm\delta^\prime$ and 0 respectively. Any
of this five models becomes nonviable if it gives rise to a tree-level
bottom mass. That is if there exists a $k$ and a $l$ for wich
$Y_{\chi_k}+Y_{D_l}=-Y_{H_\phi}$. For example, if
$Y_{H_\phi}=2\delta^\prime$ then $i=j=1$ and $k=l=2$ satisfy the
previous equations; this is signaled in Table (2) by the entry
(1,1)$_U$; (2,2)$_D$ in the Class A column and the $2\delta^\prime$ row.
The fact that in Table (2) there is at least one D-type entry for every
U-type one for all the five models of Class A, means that none of them is
viable. This fact can be easily understood by noticing that
$Y_{\chi_i}+Y_{U_j}$ changes sign under the interchange
$1\leftrightarrow 2$ in the $i,j$ indices, and that in Class A
$Y_{D_i}=Y_{U_i}$. Therefore, for a fixed $Y_{H_\phi}$,
$Y_{\chi_i}+Y_{U_j}=-(Y_{\chi_k}+Y_{D_l})$.

\noindent
\underline{CLASS B}\\
$Y_{\chi_i}=Y_{\psi_i}=\delta_i$ and $Y_{D_i}=Y_{E_i}=Y_{U_i}=\delta_i^\prime;
i=1,2,3$. There are nine different models in this class characterized by
$Y_{H_\phi}=(\delta\pm\delta^\prime),-(\delta\pm\delta^\prime),\pm\delta,
\pm\delta^\prime$ and 0 respectively. Since again $Y_{D_i}=Y_{U_i}$ none
of those models is viable.\\
\underline{CLASS C}\\
$Y_{D_i}=Y_{\psi_i}=\delta_i$ and $Y_{\chi_i}=Y_{E_i}=Y_{U_i}=\delta_i^\prime;
i=1,2,3$. There are now eleven different models in this class
characterized by
$Y_{H_\phi}=\pm 2\delta^\prime,(\delta\pm\delta^\prime),
-(\delta\pm\delta^\prime),\pm\delta,
\pm\delta^\prime$ and 0. As can be seen from Table (2) for
$\delta\neq\pm\delta^\prime,\pm 2\delta^\prime,\pm 3\delta^\prime$ and
$\delta^\prime\neq0$,
there are two models
in which only one U-type mass and none D-type one develops at
tree-level. These models are:\\
\underline{Mark I}. For a Higgs field with (U(1)$_Y$,U(1)$_H$) hypercharges
$(-1,2\delta^\prime)$.\\
\underline{Mark II}. For a Higgs field with (U(1)$_Y$,U(1)$_H$) hypercharges
$(-1,-2\delta^\prime)$.\\
The rest of the models in this classd are non-viable because a
tree-level bottom mass arises in them.\\
\underline{CLASS D}\\
This is a special class which is a particular case of Classes A,B, and
C, for which $\delta=\delta^\prime$, which in turn implies
$Y_{E_i}=Y_{\psi_I}=Y_{D_i}=Y_{\chi_i}=Y_{U_i}$. As far as the quark
mass spectrum is concerned this class is equivalent to class A.

Two comments: first, in Ref.\cite{koca}, the class of solutions A, B,
and C were all lumped together in class D, which in turn forbids the
two models classified as Mark I and Mark II above. Second, adding right-handed
neutrino fields does not change our analysis at all, either by setting
$Y_{N_1}=-Y_{N_2}=\delta, Y_{N_3}=0$ (or permutations of the indices
1,2,3); or
by imposing $Y_{N_i}=0, i=1,2,3$ in order to implement the seesaw
mechanism\cite{seesaw}.

\section{SU(3)$_c\otimes$SU(2)$_L\otimes$U(1)$_Y\otimes$U(1)$_H$ SUSY}
For the supersymmetric extension of the standard model (SSM) new particles of
spin 1/2 are introduced which are the supersymmetric partners of the Higgs
fields and gauge bosons. These higgsinos and gauginos do not contribute to
the U(1)$_{Y}$ anomaly because they are chosen vector-like  with
respect to the quantum numbers of the SM, in such a way that the SM
relationship Q$_{EM}$=T$_{3L}$+Y/2 holds, which in turn implies the U(1)$_Y$
charges in Table (3). As in the minimal SSM two different
supermultiplets of Higgs fields $\phi_{D}$ and $\phi_U$ are introduced.

Now, the simplest way to have a gauge symmetry U(1)$_H$ anomaly-free when the
new spin 1/2 members of the supermultipletes are included, is to
demand that these new fermions are vector-like with respect to U(1)$_H$.
That is, to impose Y$_{\eta_s}=\kappa$Y$_{SM_\eta}$, with $\kappa$ an
arbitrary constant. For
$\kappa=\alpha_3$, $\phi_U$ and $\phi_D$ will produce tree level masses for
the top and bottom quark respectively. This particular solution is not
consistent with (ii).

But is there other solution to the anomaly constraint equations which is
consistent with (ii)? Let us see:

If all the electrically neutral gauginos are allowed to have Majorana masses,
then Y$_{\gamma}$ = Y$_{\gamma^\prime}$ = Y$_g$ = Y$_{(W,B)}$=0.
Now, the higgsinos
$\stackrel{\sim}{\phi}_U$ and $\stackrel{\sim}{\phi}_D$ do not
carry a generational index, but if they are to produce masses at least
for the third generation, then their charges have to be related to the
charges of the third family (see the third paper in Ref.\cite{papa}).
If this is the case then the anomaly
cancellation equations are Eqs. (1)-(6) for $i=1,2$, but for $i=3$ they are:

\begin{eqnarray}
{[{\rm SU}(2)_L]^2U(1)_H}&:& Y_{\psi_3}+3Y_{\chi_3}+Y_{\phi_U}+Y_{\phi_D}=0\\
{[{\rm SU}(3)_c]^2U(1)_H} &:& 2Y_{\chi_3} + Y_{U_3}+Y_{D_3}=0\\
{[U(1)_Y]^2U(1)_H }&:& 4Y_{E_3}+\frac{2}{3}Y_{\chi_3} +
\frac{16}{3}Y_{U_3} + \frac{4}{3}Y_{D_3}
+2(Y_{\psi_3}+Y_{\phi_U}+Y_{\phi_D})= 0\\
U(1)_Y[U(1)_H]^2 &:& -Y^2_{\psi_3}+Y^2_{E_3}+Y^2_{\chi_3}
-2Y^2_{U_3}+Y^2_{D_3} +Y^2_{\phi_U}-Y^2_{\phi_D}=0\\
{[{\rm grav}]^2U(1)_H }&:& 2Y_{\psi_3} + Y_{E_3}
+2Y_{\phi_U}+2Y_{\phi_D}= 0\\
{[U(1)_H]^3 }&:& 2Y^3_{\psi_3} + Y^3_{E_3} + 6Y^3_{\chi_3}
+ 3Y^3_{U_3}+3Y^3_{D_3}+2Y^3_{\phi_U}+2Y^3_{\phi_D} = 0
\end{eqnarray}
There are two solutions to these equations. The first one is
$Y_{\chi_3}=Y_{E_3}/6=-Y_{U_3}/4=Y_{D_3}/2=-Y_{\phi_D}/3=-Y_{\psi_3}/3$
and $Y_{\phi_U}=-Y_{\phi_D}$. The second one is
$Y_{\chi_3}=Y_{E_3}/6=-Y_{U_3}/4=Y_{D_3}/2=-Y_{\phi_D}/3$ and
$Y_{\psi_3}=-Y_{\phi_U}$. For the first solution masses for the Up and
Down sector are generated simultaneously, and for the second solution
$Y_{U_3}+Y_{\chi_3}\neq -Y_{\phi_U}$, failing to give a mass for the Up
sector.

So for this extension of the SSM, the U(1)$_{H}$ anomalies can not vanish
simultaneously with the generation of only a tree-level mass for the top
quark. The alternative is to go to higher mass scales and cancell the U(1)$_H$
anomalies by a Green-Schwarz mechanism\cite{gs}. Then,
for SUSY to work there must be a $\mu$-term, meaning that the
hypercharges of the Higgs fields can be changed\cite{nir}. But this analysis
has allready been carried through in the literature\cite{papa,shrock,nir}.

\section{ACKNOWLEDGMENTS}
This work was partially supported by CONACyT in M\'exico and COLCIENCIAS
in Colombia.

\pagebreak

\underline{Table (1)}\\
U(1)$_Y$ and U(1)$_H$ charges for the known fermions. i=1,2,3 is a flavor
index related to the first, second and third generations. The Y$_{SM}$ values
stated are family independent.

\vspace{.5cm}

\begin{tabular}{||l|cccccc||}  \hline
 & $\psi_{i,L}$=(N$_i$,E$_i$)$_L$ & E$^c_{i,L}$ & $\chi_{i,L}=$
(U$_i$,D$_i$)$_L$ & U$^c_{i,L}$ & D$^c_{i,L}$ & N$^c_{i,L}$ \\ \hline
Y$_{SM}$ & $-1$ & 2 & 1/3 & $-4/3$ & 2/3 & 0 \\
Y$^i_{H_\eta}$ & Y$_{\psi_i}$ & Y$_{E_i}$ & Y$_{\chi_i}$ &
Y$_{U_i}$ & Y$_{D_i}$ & Y$_{N_i}$  \\  \hline
\end{tabular}

\vspace{.5cm}

\underline{Table(2)}\\
Summary of three-level mass term for all the possible models for the local
gauge group SU(3)$_c\otimes$SU(2)$_L\otimes$U(1)$_Y\otimes$U(1)$_H$. A Higgs
field with a hypercharge Y$_{H_\phi}$ different to the ones in the first
column does not produce a mass term in the quark sector.

\vspace{.5cm}

\hspace{-1.5cm}
\begin{tabular}{||l|c|c|c||} \hline
Y$_{H_\phi}$ & CLASS A & CLASS B & CLASS C \\ \hline
2$\delta^\prime$ & (1,1)$_U$;(2,2)$_D$ &   & (1,1)$_U$ \\
$-2\delta^\prime$ & (2,2)$_U$;(1,1)$_D$ &   & (2,2)$_U$ \\
0 & (1,2)$_U$;(2,1)$_U$;(3,3)$_U$;(1,2)$_D$;(2,1)$_D$;(3,3)$_D$ &
(3,3)$_U$;(3,3)$_D$ & (1,2)$_U$;(2,1)$_U$;(3,3)$_U$;(3,3)$_D$ \\
$\delta^\prime$ & (1,3)$_U$;(3,1)$_U$;(2,3)$_D$;(3,2)$_D$ &
(3,1)$_U$;(3,2)$_D$ & (1,3)$_U$;(3,1)$_U$;(2,3)$_D$ \\
$-\delta^\prime$ & (2,3)$_U$;(3,2)$_U$;(1,3)$_D$;(3,1)$_D$ &
(3,2)$_U$;(3,1)$_D$ & (2,3)$_U$;(3,2)$_U$;(1,3)$_D$ \\
$\delta+\delta^\prime$ &  & (1,1)$_U$;(2,2)$_D$ & (2,2)$_D$ \\
$-\delta+\delta^\prime$ &  & (2,1)$_U$;(1,2)$_D$ & (2,1)$_D$ \\
$\delta-\delta^\prime$ &  & (1,2)$_U$;(2,1)$_D$ & (1,2)$_D$ \\
$-\delta-\delta^\prime$ &  & (2,2)$_U$;(1,1)$_D$ & (1,1)$_D$ \\
$\delta$ &  & (1,3)$_U$;(2,3)$_D$ & (3,1)$_D$ \\
$-\delta$ &  & (2,3)$_U$;(1,3)$_D$ & (3,2)$_D$ \\ \hline
\end{tabular}

\vspace{0.5cm}

\underline{Table (3)}\\
U(1)$_Y$ and U(1)$_{H}$ charges for the sparticles of spin 1/2.
$\stackrel{\sim}{\gamma}$ and $\stackrel{\sim}{\gamma^\prime}$ are the
gauginos related to U(1)$_Y$ and U(1)$_H$
respectively, \~g stand for the eight gluinos, etc.

\vspace{.5cm}

\begin{tabular}{||l|cccccc||} \hline
 & $\stackrel{\sim}{\phi}_U$ & $\stackrel{\sim}{\phi}_D$ &
$\stackrel{\sim}{\gamma}$ & $\stackrel{\sim}{\gamma}^\prime$ & \~g & (\~W,\~B)
\\ \hline
Y$_{SM}$ & 1 & $-1$ & 0 & 0 & 0 & 0 \\
Y$_{\eta_s}$ & Y$_{\phi_U}$ & Y$_{\phi_D}$ & Y$_\gamma$ & Y$_{\gamma^\prime}$ &
Y$_g$ & Y$_{(W,B)}$ \\ \hline
\end{tabular}

\pagebreak


\begin{thebibliography}{99}
\bibitem{hsh}
R.Barbieri and D.V.Nanopoulos, Phys. Lett. {\bf 91B}, 369 (1980); {\bf 95B},
43 (1980).

\bibitem{dssm}
L.Iba\~nez and G.G.Ross, Phys. Lett. {\bf 110B}, 215 (1982); L.Alvarez
Gaum\'e, J.Polchinki and M.Wise, Nucl. Phys. {\bf B221}, 495 (1983).

\bibitem{mhsh}
W.A.Ponce, A.Zepeda, A.H.Galeana and R.Mart\'\i nez, Phys. Rev {\bf D44}, 2166
(1991).

\bibitem{anom}
S.L.Adler, Phys. Rev. {\bf 177}, 2426 (1969); J.S.Bell and R.Jackiw, Nuovo
Cimento {\bf 51}, 47 (1969).

\bibitem{del}
A.Salam and R.Delburgo, Phys. Lett. {\bf 40B}, 381 (1972); L.Alvarez Gaum\'e
and E.Witten, Nucl. Phys. {\bf B234}, 269 (1983)

\bibitem{gs}
M.Green and J.Schwarz, Phys. Lett. {\bf 149B}, 117, (1984).

\bibitem{papa}
L.Iba\~nez and G.G.Ross, Phys. Lett {\bf 332B}, 100 (1994); E.Papageorgiu,
Z.Physik {\bf C64}, 509 (1994); P.Bin\'etruy and P.Ramond: ``Yukawa Textures
and Anomalies", LPTHE-ORSAY 94/115, UFIFT-HEP-94-19 preprint, unpublished.

\bibitem{koca}
A.Davidson, M.Koca and K.C.Wali, Phys. Rev. Lett. {\bf 43}, 92 (1979);
Phys. Rev. {\bf D20}, 1195 (1979); {\bf D21}, 787 (1980).

\bibitem{shrock}
V.Jain and R.Shrock: ``{\it Models of Fermion Mass Matrices based on Flavor
and Generation-Dependent U(1) gauge symmetry}", ITP-SB-94-55 preprint,
unpublished.

\bibitem{nir}
Yosef Nir: ``{\it Gauge Unification, Yukawa hierarchy and the $\mu$ problem}",
hep-ph/9504312, Wis-95/18/Apr-PH preprint, unpublished.

\bibitem{po}
W.A.Ponce, Phys. Rev. {\bf D36}, 962 (1987).

\bibitem{seesaw}
M.Gell-Mann, P.Ramond, and R.Slansky, in ``{\it Supergravity}", proceedings of
the workshop, Stony Brook, N.Y. (1979), edited by P.van Nieuwenhuizen and
D.Z.Freedman (North-Holland, Amsterdam 1979). p.315; T.Yanahida, in  ``{\it
Proceedings of the workshop on Unified Theories and the Baryon Number in the
Universe}", edited by A.Sawada and A. Sugamoto (KEK report No. 79-18,
Tsukuba-Gun, Ibaraki-Ken, Japan, 1979).

\end{thebibliography}
\end{document}